\documentclass[aps,prb,preprint,floatfix]{revtex4-1}

\usepackage{graphicx}
\usepackage{amsmath}
\usepackage{amsfonts}
\usepackage{amssymb}
\usepackage{dcolumn}
\usepackage{braket}
\usepackage{commath}
\usepackage{epstopdf}
\usepackage{hyperref}

\usepackage{color}

\usepackage{ulem}   % to strike things out
\usepackage[utf8]{inputenc}
\usepackage[english]{babel}
\usepackage[dvipsnames]{xcolor}
\usepackage{subfig}
\usepackage{physics}
\usepackage{amsmath}
\usepackage{amsfonts}
\usepackage{amssymb}
\usepackage{graphicx}
\usepackage{bm}
\usepackage{wrapfig}
\usepackage[left=2cm,right=2cm,top=2cm,bottom=2cm]{geometry}
\usepackage{subfig}
\usepackage{float}
\restylefloat{table}
\usepackage{listings}
\usepackage{color}

\begin{document}

\title{Numerical and Analytical Study of the Bound States of the  $-\alpha/x^2$ Potential}

\author{Thanh Xuan Nguyen}

\author{F.~Marsiglio}
\email{fm3@ualberta.ca}

\affiliation{Department of Physics, University of Alberta, Edmonton, Alberta, Canada, T6G~2E1}

\begin{abstract}
The quantum mechanical bound states of the $-{\alpha}/x^2$ potential are truly anomalous. We revisit this problem by adopting a slightly
modified version of this potential, one that adopts a cutoff in the potential arbitrarily close to the origin. The resulting solutions are completely
well-defined and ``normal.'' We present results here as a case study in undergraduate research --- two independent methodologies are used: one
analytical (with very unfamiliar non-elementary functions) and one numerical (with very straightforward methodology). These play complementary
roles in arriving at solutions and achieving insights in this problem.
\end{abstract}

\date{\today }
\maketitle

\section{introduction}

In this paper we wish to study the quantum mechanics of a particle subject to the potential $-{\alpha}/x^2$ %(with $\tilde{\alpha} \equiv \alpha \hbar^2/(2m_0) > 0$ 
(with $\alpha > 0$ and $m_0$ the mass of the particle) in the domain $0<x< \infty$. This potential defies
our intuition and expectations, even for the ``quantum world.'' A number of treatments exist already in the literature,\cite{case50,morse53,landau77,coon02,essin06,griffiths18}
where the difficulties connected with this potential are worked through and discussed. In particular, these references point out that for 
$2m_0 \alpha/\hbar^2 \equiv \rho_0^2 < 1/4$ there are no bound states,  while for $\rho_0^2  > 1/4$, there are an
infinite number of bound states, with energies of arbitrarily negative value. In Refs.~[\onlinecite{essin06}] and [\onlinecite{griffiths18}] a 
``regularization'' procedure is used to restore ``proper'' quantum mechanical solutions to the problem; this consists of a displaced ``wall'' so that the origin 
(and hence the singular behavior of the potential) is no longer accessible. A different approach, which we will adopt below, is suggested in 
Ref.~[\onlinecite{landau77}] in their approximate analytical treatment near the origin. 
In this case, the potential is made to be a constant below some small value of $x$, so that the potential is continuous for $x>0$.

Why present another study of the solutions for this particular potential? First, as we will note below, this problem is not as unphysical as one might first 
think. It shows up immediately in the study of an electron binding to a polar molecule,\cite{levy-leblond67} and also arises
naturally in problems with cylindrical geometry, as occurs, for example, in the problem of a charged particle in a magnetic field, when adopting the symmetric gauge.
Secondly, while this problem has an analytical solution, not only is the solution given in terms of non-elementary functions (modified Bessel function), but these functions
are of \textit{imaginary} order. While undergraduate students can now access these functions through a variety of packages generally available to them, this process
remains very ``black-box-like'' and is perhaps of limited use to the typical student. Instead, here we adopt a numerical matrix method,\cite{marsiglio09,jelic12,randles19} which students
can implement on their own, and this allows them to explore these solutions and confirm the validity of the analytical solutions (an intriguing inversion of the usual validation
process!). This method requires mathematical knowledge at the undergraduate first year level only, but does require software to diagonalize large matrices. Most importantly, the machinery
required is generally suited to problems with ordinary binding potentials, i.e. it is not specifically for this (somewhat strange) particular problem, but can readily be
applied to it.

As suggested in the previous paragraph, the problem of the behavior of a particle in the $-\alpha/x^2$ ($\alpha > 0$) 
potential in the domain $0<x< \infty$ can be viewed in several ways. First, it can be thought of
as a one-dimensional problem, ``conjured'' up to illustrate various pathologies. Secondly, the variable `$x$' can be viewed as the radial coordinate in a problem with spherical
symmetry, where the one-body potential is given by $-\alpha^\prime/x^2$ so that $-{\alpha}/x^2 \equiv -\bigl(\alpha^\prime - [(\hbar^2/2m_0) \ell (\ell + 1)]\bigr)/x^2$ represents the \textit{effective} attractive potential governed by an inverse square law. In this case, the requirement 
that $x>0$ naturally arises because the radial coordinate is by definition non-negative.
Finally, if we imagine a problem with a potential with cylindrical symmetry, i.e. one that is independent of $z$ and dependent only on the polar coordinate $r$, where the polar
coordinates $(r,\theta)$ are defined through $x=r{\rm cos} \theta$, $y=r{\rm sin} \theta$, then it is prudent to rewrite the three-dimensional time-independent 
Schr\"odinger Equation in polar coordinates, with $\psi = \psi(r,\theta,z)$. Using separation of variables, $\psi \equiv R(r)  \Theta(\theta)  Z(z)$ leads to the 
$z$-dependence which is a plane-wave solution, the $\theta$-dependence is given simply by $~e^{i\ell \theta}$, with $\ell$ an integer, and the $r$-dependence will be 
governed by a differential equation. Following the procedure in three dimensions, where we introduced an auxiliary radial wave function defined by $u(r) \equiv rR(r)$,
we similarly adopt the auxiliary wave function defined by
\begin{equation}
u(r) \equiv \sqrt{r} R(r) \ \ \ \ \ \ \ (2D)
\label{aux_wave}
\end{equation}
that can be shown to satisfy
\begin{equation}
-{\hbar^2 \over 2m_0} {d^2u(r) \over dr^2} + V_{\rm eff}(r) u(r) = E u(r),
\label{2d_eq}
\end{equation}
where
\begin{equation}
V_{\rm eff}(r) \equiv V(r) + {\hbar^2 \over 2m_0} \bigl( \ell^2 - {1 \over 4} \bigr){1 \over r^2}.
\label{veff}
\end{equation}
Remarkably, before even discussing the form of the one-body potential $V(r)$, the $1/r^2$ attractive potential already appears in this two-dimensional problem
(provided $\ell = 0$). Even more interesting, the value of the coefficient is precisely at the demarcation of the peculiar behavior noted above and in the references.

\section{THE FORMALISM}

\begin{figure}[H]
\centering  
\includegraphics[scale=.5]{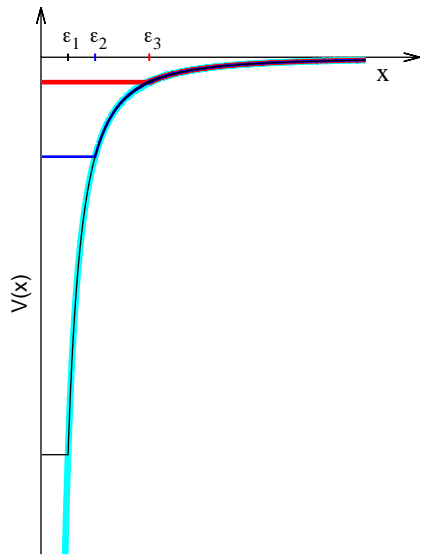}
 \caption
{Sketch of the truncated potential as defined in Eq.~(\ref{pot}): $V(x) = -\alpha/x^2$ for $x > \epsilon$, and $V(x) = {\rm constant}$ for $0<x< \epsilon$. Continuity of the
potential requires that the constant $ = -\alpha/\epsilon^2$. Here a variety of choices for the cutoff is depicted (with $\epsilon_1 < \epsilon_2 < \epsilon_3$) depicted
with black, blue, and red curves, respectively. The underlying $-\alpha/x^2$ potential is shown with a thick light blue curve.
}
\label{fig1}
%\label{Regularized Potential plot}
\end{figure}

We wish to solve the one-dimensional Schr\"odinger equation,
\begin{equation}
-{\hbar ^2 \over 2 m_0} {d^2 \psi (x) \over dx^2} + V(x) \psi (x)= E \psi (x),
\label{schro}
\end{equation}
where $V(x)$ is specified by
\begin{align}
V_{\epsilon}(x) =\begin{cases}
- \dfrac{\alpha}{{\epsilon}^2} \quad & \text {if $0 < x < \epsilon$}
\\
-\dfrac{\alpha}{x^2} \quad & \text {if $\epsilon < x < \infty$}
\end{cases}
\label{pot}
\end{align}
with a cutoff near the origin (at $x=\epsilon$) that avoids the singularity that causes the problems. This potential is sketched in Fig.~\ref{fig1}.
The strategy is to solve this problem (which has no difficulties),
and allow $\epsilon \rightarrow 0$ so that we can try to track the problems as they arise. Equation~(\ref{schro}) with $V_{\rm eff}$ given by Eq.~(\ref{pot}) is
precisely the kind of problem that was tackled in Ref.~[\onlinecite{jugdutt13}] through matrix mechanics, and we will follow the procedure outlined there. In addition, it is straightforward (but not for undergraduates!) to provide an analytical solution, and we will first proceed in this way.
%figure 1

\subsection{Analytical Solution}

To solve the Schr\"odinger equation we first divide the domain into the two regions.
For $0 < x < \epsilon$, the equation is
\begin{equation}
- \dfrac{\hbar ^2}{2 m_0} \dfrac{d^2 \psi (x)}{dx^2} - \dfrac{\alpha}{{\epsilon}^2} \psi (x) = E \psi (x),
\label{schro1}
\end{equation}
with solution
\begin{equation}
\psi _1 (x) = A \sin \left( \sqrt{\dfrac{2 m_0}{\hbar ^2}\left(E+{\alpha \over \epsilon^2}\right)}x \right) = A\sin\left(qx\right),
\label{soln1}
\end{equation}
where $q \equiv \sqrt{2 m_0\left(E+{\alpha/\epsilon^2}\right)/\hbar^2}$ and we have dropped the $\rm{cos} (qx)$ solution to ensure the proper behaviour at the origin. 
For $x \geq \epsilon$ % \leqslant x$,
we have
\begin{equation}
- \dfrac{\hbar ^2}{2 m_0} \dfrac{d^2 \psi (x)}{dx^2} - \dfrac{\alpha}{{x}^2} \psi (x) = E \psi (x).
\label{schro2}
\end{equation}
Upon substituting $\rho_0^2 \equiv 2m_0 \alpha /\hbar^2$ and $\rho \equiv \kappa x$, where $\kappa^2 \equiv -2m_0 E /\hbar^2$ this becomes
\begin{equation}
{d^2\psi_2(x) \over d\rho^2} -\left( \rho^2 - \rho_0^2 \right)\psi_2(x) = 0.
\label{schro2b}
\end{equation}
Using $\psi_2 \equiv \sqrt{\rho} \phi_2(\rho)$, we obtain
\begin{equation}
\rho^2 {d^2\phi(\rho) \over d\rho^2} + \rho {d\phi(\rho) \over d\rho} - \left(\rho^2 + \nu^2 \right) \phi(\rho),
\label{schro2c}
\end{equation}
where $\nu \equiv ig \equiv i\sqrt{\rho^2_0 - {1 \over 4}}$ is pure imaginary for $\rho_0 > 1/2$. Equation~(\ref{schro2c}) is just the Bessel equation with solutions given by a linear
combination of the modified Bessel functions $K_{\nu}(\rho)$ and $I_{\nu}(\rho)$, with imaginary index given by $ig$ when $\rho_0^2 > 1/4$.
The $I_{\nu}(\rho)$ solutions diverge as $\rho$ increases, so we retain only the $K$ solution. Therefore, the solution to the original problem is
\begin{equation}
\psi_2(x) = B \sqrt{\kappa x} K_{ig}(\kappa x).
\label{schro2d}
\end{equation}

The eigenvalues $E_n(\epsilon)$ are determined by matching the wave functions and their derivatives at $x = \epsilon$,
\begin{eqnarray}
\psi_1(\epsilon) &=& \psi_2(\epsilon)
\label{boundary1} \\ 
\nonumber \\
\left. {d\psi_1(\rho) \over d \rho} \right|_{\kappa \epsilon} &=& \left. {d\psi_2(\rho) \over d \rho} \right|_{\kappa \epsilon}.
\label{boundary2}
\end{eqnarray}
The condition to determine the energy $E_n(\epsilon)$ is therefore
\begin{equation}
{1 \over q \epsilon} {\rm tan}(q\epsilon) = {2K_{ig}(\rho_\epsilon) \over K_{ig}(\rho_\epsilon) + 2 \rho_\epsilon {dK_{ig}(\rho) \over d\rho}|_{\rho = \rho_\epsilon}},
\label{energy}
\end{equation}
where $\rho_\epsilon \equiv \kappa \epsilon$. Since $q\epsilon = \sqrt{\rho_0^2 - \rho_\epsilon^2}$, and $g$ is a function of $\rho_0^2$ only, this means that we
seek a solution, $\rho_\epsilon^2 = f(\rho_0^2)$, where $f$ is some function. The important point is that the solution, $\rho_{\epsilon}$, depends only on $\rho_0^2$, and
does not depend on $\epsilon$. 
So, recalling the definition of $\kappa$, we have
\begin{equation}
E = -{\alpha \over \epsilon^2}{f(\rho_0^2) \over \rho_0^2}.
\label{energyb}
\end{equation}
Another way of writing this in dimensionless units is
\begin{equation}
{E  \over \alpha/\epsilon^2} = -{f(\rho_0^2) \over \rho_0^2}.
\label{energyc}
\end{equation}
Equation~(\ref{energy}) needs to be solved for the eigenvalues for a given $\rho_0^2$ and $\epsilon$. Equation~(\ref{energyb}) tells us that the $\epsilon$ dependence  is
remarkably simple, and the energy simply goes as  $\approx 1/\epsilon^2$. Thus the bound state energies {\it all} diverge as $\epsilon \rightarrow 0$. Less obvious is 
how many bound state solutions ($E < 0$) exist. We will find that, like the Coulomb potential there exist an infinite number, even with the cutoff provided by a finite $\epsilon$.
Once an eigenvalue is determined then either of the conditions given by 
Eqs.~(\ref{boundary1}) or (\ref{boundary2}) determines the coefficient $B$ in terms of $A$.
Finally, normalization of the wave function determines the remaining coefficient. These equations are simply solved,\cite{remark} and the solutions will be displayed
alongside the numerical ones. Before showing these we discuss the numerical solution.

\subsection{Numerical Solution}

Following Refs.~[\onlinecite{marsiglio09}] and [\onlinecite{jugdutt13}], we embed the potential given in Eq.~(\ref{pot}) in an infinite square well extending from $0 < x < a$, where the width
$a>>\epsilon$ is taken to be large enough to obtain accurate results for at least the low-lying energy levels and their eigenstates. A reasonable value of $a$ requires some
experimentation and has to be coordinated with a reasonable choice for a cutoff in the number of basis states (since we can't work with an infinite number of these).
Then we can expand the wave function in a basis set consisting of 
\begin{equation}
\phi_n(x) = \sqrt{2 \over a} {\rm sin}\left({n \pi x \over a}\right),
\label{basis_set}
\end{equation}
i.e. 
\begin{equation}
\psi(x) = \sum_{n=1}^\infty c_n \phi_n(x),
\label{expansion}
\end{equation}
and we arrive at the matrix equation,
\begin{equation}
\sum_{m=1}^{N_{\rm max}} H_{nm} c_m = E c_n,
\label{matrix}
\end{equation}
where $N_{\rm max}$ is a cutoff, controlled to give converged results. The matrix elements are given by
%
%\begin{figure}[H]
%\centering  
%\includegraphics[scale=.4]{photos/The_potential_wider_a}
% \caption
%{ 
%The regularized potential is placed inside infinite square wells with the width $a$.
%\label{wider widths a to describe higher states}
%} 
%\end{figure}
%
\begin{equation}
H_{nm} = H^{K}_{nm} + H^{V}_{nm},
\label{hamtot}
\end{equation}
where the kinetic contribution is diagonal,
\begin{equation}
H^{K}_{nm} = \delta_{nm} {\hbar^2 \pi^2 n^2 \over 2m_0 a^2},
\label{hkin}
\end{equation}
and the potential energy contribution requires integration over the two regions defined in Eq.~(\ref{pot}) (with the 2nd region truncated at $x=a$):
\begin{equation}
H^V _{nm} = -{2 \alpha \over a \epsilon^2} \left[ \int ^{\epsilon}_0 dx \sin{\left( \dfrac{n \pi x}{a} \right)} \sin{\left( \dfrac{m \pi x}{a}\right)}
+ \int ^{a}_{\epsilon} dx \dfrac{\epsilon^2}{x^2} \sin{\left( \dfrac{n \pi x}{a} \right)} \sin{\left( \dfrac{m \pi x}{a}\right)} \right].
\end{equation}
This expression simplifies to
\begin{eqnarray}
H^V _{nm} = -{\alpha \over a^2} &&\Biggl\lbrace \ {a \over \epsilon} \left[ \delta_{nm} +  (1-\delta_{nm}) {\rm Sinc} \left((n-m)\pi {\epsilon \over a} \right) - 
{\rm Sinc} \left((n+m)\pi {\epsilon \over a} \right)
\right] \phantom{\Biggr\rbrace} \nonumber \\
\phantom{\Biggl\lbrace} &&+ \ L_2\left(n+m,{\epsilon \over a}\right) -  L_2\left(n-m,{\epsilon \over a}\right) \Biggr\rbrace,
\end{eqnarray}
where ${\rm Sinc}(\rho) \equiv {\rm sin}(\rho)/\rho$ and
\begin{equation}
L_2(n,\rho) \equiv \int_\rho^1 dy {1 \over y^2} \lbrack 1 - {\rm cos}(n\pi y) \rbrack
\label{l2}
\end{equation}
can be evaluated numerically or rewritten in terms of the Sine Integral,\cite{abramowitz64} ${\rm Si}(z)$. In practice, we rewrite Eq.~(\ref{matrix}) in dimensionless
form by dividing both sides by $E_0 \equiv \hbar^2 \pi^2/(2m_0 a^2)$ and therefore find the eigenvalues in units of $E_0$. 
The dimensionless matrix elements are
\begin{eqnarray}
h_{nm} \equiv {H_{nm} \over E_0} = n^2\delta_{nm} - {\rho_0^2 \over \pi^2} &&\Biggl\lbrace \ {a \over \epsilon} \left[ \delta_{nm} +  (1-\delta_{nm}) {\rm Sinc} \left((n-m)\pi {\epsilon \over a} \right) - 
{\rm Sinc} \left((n+m)\pi {\epsilon \over a} \right)
\right] \phantom{\Biggr\rbrace} \nonumber \\
\phantom{\Biggl\lbrace} &&+ \ L_2\left(n+m,{\epsilon \over a}\right) -  L_2\left(n-m,{\epsilon \over a}\right) \Biggr\rbrace.
\end{eqnarray}
The matrix diagonalization is now completely determined by these numbers, once $\rho_0$, $\epsilon/a$, and $N_{\rm max}$ are specified. Recall that $\rho_0 > 1/2$
ensures that there are bound states, and we want to take $\epsilon/a$ closer and closer to zero.

\section{Results and Discussion}

For $\rho_0 \equiv \sqrt{2m_0 \alpha/\hbar^2} < 1/2$ (including negative values) there are no bound states, i.e. states with energy less than zero. We have confirmed this numerically. In this paper we focus on the regime where
there are definite bound states. 

\subsection{The bound state energies}

\begin{figure}[H]
\centering  
\includegraphics[scale=0.50]{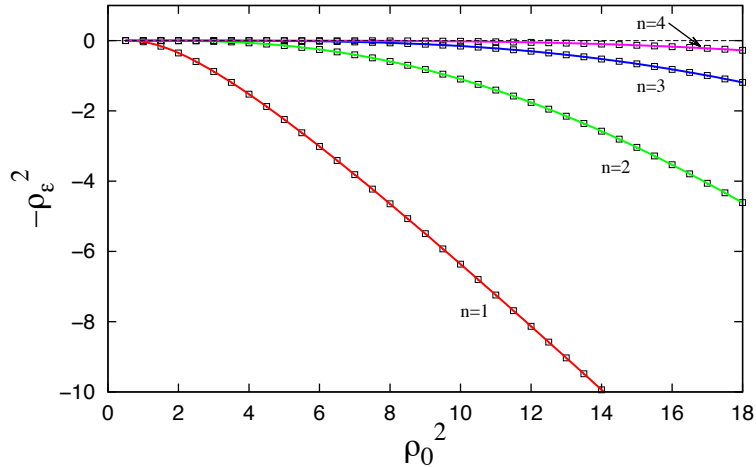}
 \caption
{ 
Four lowest eigenvalues, $\rho_\epsilon^2 \equiv -E_n\rho_0^2 \epsilon_2/\alpha$, for $n=1,2,3,4$, as a function of the strength of the potential, $\rho_0^2 \equiv 2m_0 \alpha/\hbar^2$. Note that these energies have been computed for the potential with a cutoff [see Eq.~(\ref{pot})], but, written in these units, the results are independent of the cutoff position, $\epsilon$, as was the case in Ref.~[\onlinecite{essin06}] with their different regularization procedure. Symbols denote the analytical results obtained by solving Eq.~(\ref{energy}), and the curves denote
the numerical results achieved by exact diagonalization of $4900 \times 4900$ matrices, as discussed in the previous section. These latter results become inaccurate as the bound state energies approach zero, as is expected since the wave function becomes more extended in this case and they begin to ``feel'' the effects of the wall of the infinite square well potential used to define the basis set. Note that the analytical solution indicates that an infinite number of bound states occur for any given potential strength, no matter how
small, as long as $\rho_0^2 > 1/4$. The numerical results require an actual choice of $\epsilon$, and we used $\epsilon/a = 0.001$. For the numerical results, eventually the
higher `n' excited states (not shown) become ``unbound'' due to the presence of the infinite square well, and will disagree with the analytical results.
}
\label{fig2}
%\label{a typical result}
\end{figure}

\begin{figure}[H]
\centering  
\includegraphics[scale=.70]{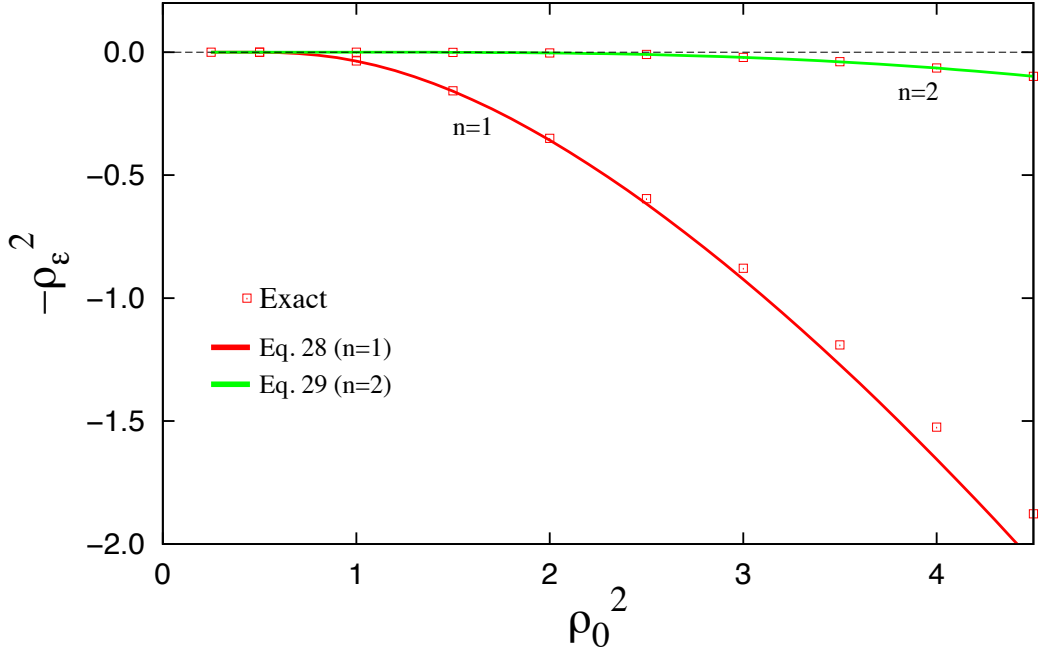}
 \caption
{ 
The two lowest eigenvalues, $\rho_\epsilon^2 \equiv -E_n\rho_0^2 \epsilon^2/\alpha$, for $n=1,2$ (same as in Fig.{\ref{fig2}}), as a function of the strength of the 
potential, $\rho_0^2 \equiv 2m_0 \alpha/\hbar^2$, but over a more limited range. The exact results are shown (points) along with the approximate result given by
Eqs.~(\ref{ener1}) and (\ref{enern}) (curves) for the two lowest energy eigenstates. The agreement with the higher excited states is even more accurate, 
but these energies are very close to zero.}
 \label{fig3}
%\label{a typical result}
\end{figure}

In Fig.~(\ref{fig2}) we show the exact results for the first 4 bound states as a function of $\rho_0^2$.
%figure 2
In fact, every strength of potential shown supports an infinite number of bound states, but these very quickly become very weakly bound with increasing quantum number, $n$.
This is seen analytically, by taking the expression for $K_{ig}(x)$ with small argument (i.e. energy close to zero):
\begin{equation}
K_{ig}(x) \approx -\sqrt{2 \pi g e^{-\pi g} \over 1 - e^{-2 \pi g}}{1 \over g} {\rm sin}\left[ g \ {\ell n}({x \over 2}) - \phi(k=0)\right],
\label{smallx}
\end{equation}
where $\phi(k)$ is the argument of the Gamma function given by
\begin{equation}
\phi(k) \equiv {\rm arg} [\Gamma(1 + k + ig)] = g\psi(1+k) + \sum_{n=0}^\infty \left( {g \over 1 + k + n} - {\rm tan}^{-1}\left({g \over 1 + k + n}\right) \right)
\label{arg}
\end{equation}
and $\psi(x)$ is the Digamma function. We need $\psi(1) = -\gamma \approx -0.5772$ where $\gamma$ is Euler's constant. Inserting this into Eq.~(\ref{energy}) we find
a ground state energy given by
\begin{equation}
E_1 \rho_0^2 {\epsilon^2 \over \alpha} = -4 \ {\rm exp}\Biggl({2 \over g} \bigl[\phi(k=0) + {\rm tan}^{-1}\biggl( {g {{\rm tan}\rho_0 \over \rho_0} \over 1 - {{\rm tan}\rho_0 \over 2\rho_0}} \biggr) \bigr] \Biggr))
\label{ener1}
\end{equation}
with excited (bound) state energies given by
\begin{equation}
E_n = E_1 {\rm exp}\bigl[-{2\pi (n-1)/g} \bigr], \ \ \ \ \ \ n=1,2,3....
\label{enern}
\end{equation}
Care is required in Eq.~(\ref{ener1}) as the correct branch of the inverse tangent function is required. In Fig.~\ref{fig3} we show 
the two lowest bound state energies from Fig.~\ref{fig2}, but over a smaller range of $\rho_0^2$, alongside with the approximate results given by 
Eqs.~(\ref{ener1}) and (\ref{enern}). Agreement is very good for the ground state all the way up to $\rho_0^2 \approx 3$, even more so for $n=2$, 
and gets better for the other bound state energies (there are an infinite number of them!), which on this scale are essentially indistinguishable from zero.
In this and in subsequent figures with numerical results, we have used $4900 \times 4900$ matrices to assure convergence as a function of basis
size. In fact in most cases convergence was attained with $400 \times 400$ matrices.
%figure 3

\subsection{The bound state wave functions}

\begin{figure}[H] %[tp]
\begin{center}
\includegraphics[scale=.6]{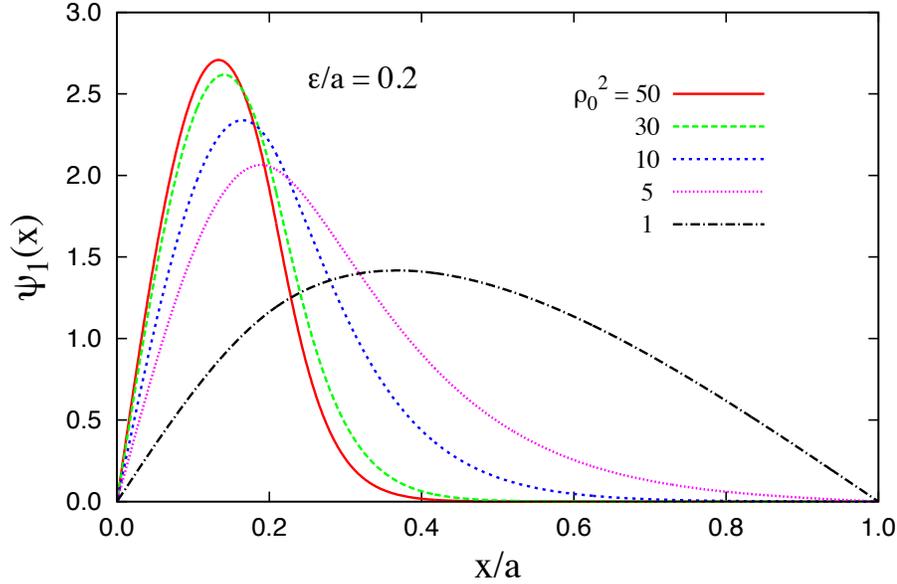}
\end{center}
\caption
{ 
The ground state wave function computed for various values of the potential strength, with the cutoff fixed at $\epsilon/a = 0.2$, and the potential ``cut off'' at large
$x$ by the presence of an infinite square well as described in the text. Here it is visually obvious that the wave function becomes more concentrated at lower values
of $x$ as the strength $\rho_0^2$ increases. It is also clear that as the strength decreases (e.g. $\rho_0^2 = 1$ or even $5$), the infinite square well is playing a
role in determining the wave function (and the bound state energy), since the wave function is significantly {\it nonzero} at the boundary ($x=a$). So the dashed curve
representing the result for $\rho_0^2 = 1$ is clearly {\it not} representative of the potential we wish to study ($-\alpha/x^2$ with a cutoff at $x=\epsilon$) because it would
like to be more extended (and therefore needs a wider infinite square well, i.e. larger value of $a$). Therefore we should use a lower value of $\epsilon/a$ if we 
wanted to know more about the results for this potential strength. This case and comparisons to the analytical results will be shown below.
} 
%\label{v0 decreases ground-state wave functions extended}
\label{fig4}
\end{figure}

\begin{figure}[H]
\centering  
\includegraphics[scale=.6]{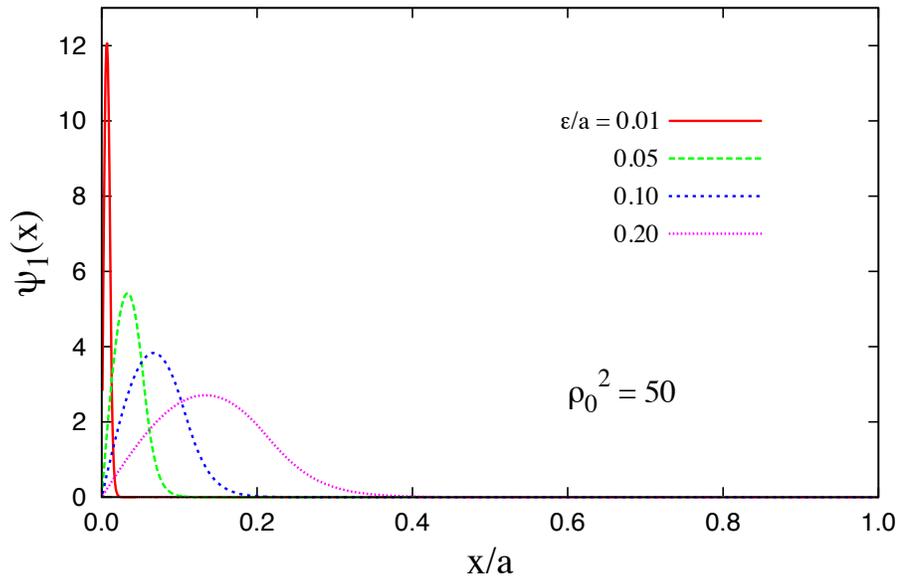}
 \caption
{ 
The ground state wave function computed for various values of the cutoff, $\epsilon/a$, for a given potential strength, $\rho_0^2 = 50$. Because this is a particularly
high value, the results are converged for all values of $\epsilon$, given this width of infinite potential well, $a$. Here it is even more visually obvious that the wave function becomes more concentrated as the value of $\epsilon/a$ is decreased. In fact, compared to the previous figure, this result makes it clear that something anomalous
is happening to the wave function as $\epsilon \rightarrow 0$, insofar as the wave function becomes very concentrated as this limiting process occurs. In contrast, in
the previous figure, the wave function was converging to a fixed result as $\rho_0^2 \rightarrow \infty$.
} 
\label{fig5}
%\label{localized wave function as epsilon decreased}
\end{figure}

Wave functions are also readily accessible. In Fig.~\ref{fig4} we show the ground state wave function obtained from the numerical approach 
(these require the eigenvector)\cite{marsiglio09} for increasing values of the strength of the potential (fixed cutoff, $\epsilon$) and in Fig.~\ref{fig5}
we plot the same function for various values of the cutoff in the potential, $\epsilon$ (fixed strength, $\alpha$, or $\rho_0^2$). 
%figure 4
%\begin{figure}[H]
%\centering  
%figure 5

\begin{figure}[H]
\centering  
\includegraphics[scale=.6]{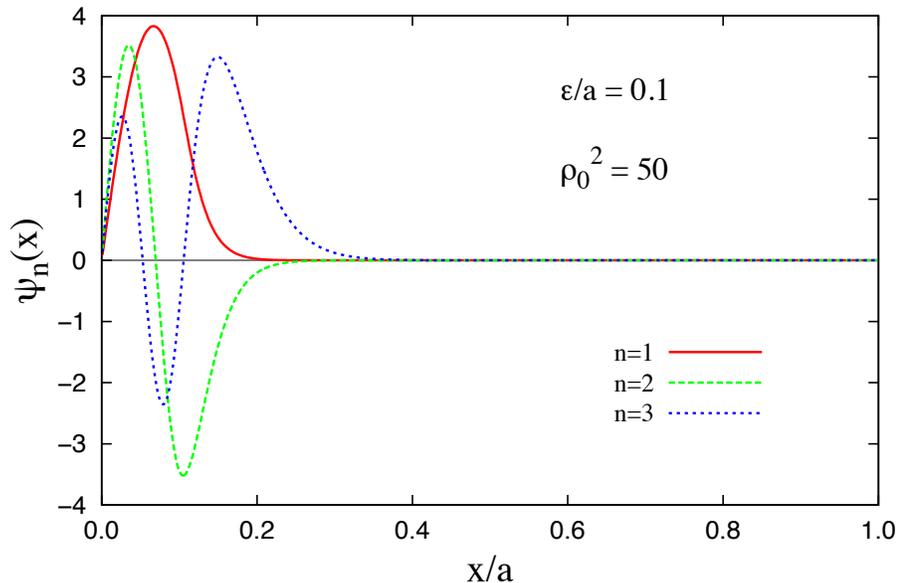}
 \caption
{ 
The first two excited wave functions, $n=2$ (green dashed curve) and $n=3$ (blue dotted curve), along with the ground state ($n=1$, red solid curve) vs. $x/a$
for $\epsilon /a = 0.1$ and $\rho_0^2 = 50$. These results were obtained numerically, and therefore with an embedding infinite square well potential, but
this embedding potential does not play a role, as evidenced by the near-zero wave function amplitude at $x=a$.
} 
\label{fig6}
\end{figure}

In either case, as we raise the potential strength or 
lower the cutoff distance, we obtain the expected behaviour, which is a movement of the wave function towards the origin. Lowering the value of the position
cutoff has a far more potent effect, because it is through this process that the problem becomes (eventually) ill-defined. These figures do illustrate, however, that with
a cutoff in the potential, the results are perfectly reasonable, i.e. non nodes in the ground state. In Fig.~\ref{fig6} we show the 
first and second excited states for certain parameter values, and they have the
standard features (one, and two nodes, respectively, zero at the origin) expected in such a problem.
%figure 6
Moreover, they are well converged, in the sense that they clearly are oblivious to the presence of the infinite square well with wall at $x=a$.
A repeat of Fig.~\ref{fig6} with a smaller value of $\epsilon$ will give a similar result, with wave functions confined more closely to the origin. However, by use of a judicious scaling
we can provide universal results. In fact we stumbled upon this through the numerical results, but a closer examination of the analytical answer shows that the wave function
can be written as
\begin{align}
\psi_n(x) = {1 \over \sqrt{\epsilon}} {1 \over \sqrt{h_n(\rho_0)}} \begin{cases}
{ {\rm sin}\bigl(\sqrt{\rho_0^2 - \rho_{\epsilon}^2} {x/\epsilon}\bigr) \over {\rm sin}\bigl(\sqrt{\rho_0^2 - \rho_{\epsilon}^2} \bigr)}  \quad & \text {if $0 < x < \epsilon$}
\\
\sqrt{x \over \epsilon} {K_{ig}(\rho_\epsilon {x/\epsilon}) \over K_{ig}(\rho_\epsilon) } \quad & \text {if $\epsilon < x < \infty$},
\end{cases}
\label{wavefn}
\end{align}
where the subscript $n$ is the quantum number implicit in the solutions for the eigenvalue tabulated by $\rho_\epsilon$, and previously shown in 
Fig.~\ref{fig2} or Fig.~\ref{fig3}. The function $h_n(\rho_0)$ is determined by normalization:
\begin{equation}
h_n(\rho_0) = \int_0^1 dy { {\rm sin}^2\bigl(\sqrt{\rho_0^2 - \rho_{\epsilon}^2} {y}\bigr) \over {\rm sin}^2\bigl(\sqrt{\rho_0^2 - \rho_{\epsilon}^2} \bigr)} 
+ \int_1^\infty dy y  {K^2_{ig}(\rho_\epsilon {y}) \over K^2_{ig}(\rho_\epsilon) },
\label{normalization}
\end{equation}
and $h_n(\rho_0)$ is written as a function of $\rho_0$ only, because (i) $\rho_\epsilon$ is a function of $\rho_0$ only, and (ii) explicit dependence on $\epsilon$ has dropped out
[see the discussion above Eq.~(\ref{energyb})]. Thus $\sqrt{\epsilon} \psi(x/\epsilon)$ is a universal (as far as $\epsilon$ is concerned) function of $x/\epsilon$.

\begin{figure}[H]
\centering  
\includegraphics[scale=.5]{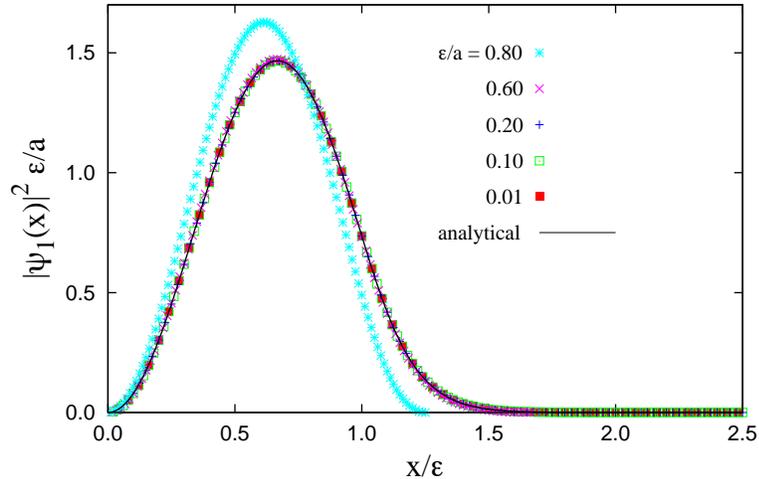}
 \caption
{ 
The probability, multiplied by $\epsilon/a$, vs. $x/\epsilon$ for a variety of values of $\epsilon/a$, obtained using the numerical matrix method. These
quickly converge to the result shown. Also included is the analytical result, shown with a black curve, in clear agreement with the numerical data. In this
case the quantity $\epsilon |\psi_1(x)|^2$ is plotted (there is no `$a$' !) vs. $x/\epsilon$, and there is only the one (universal) result for any $\epsilon$.
We used $\rho_0^2 = 50$.
} 
\label{fig7}
%\label{the general pattern of ground-state wavefunction}
\end{figure}

\begin{figure}[H]
\centering  
\includegraphics[scale=.5]{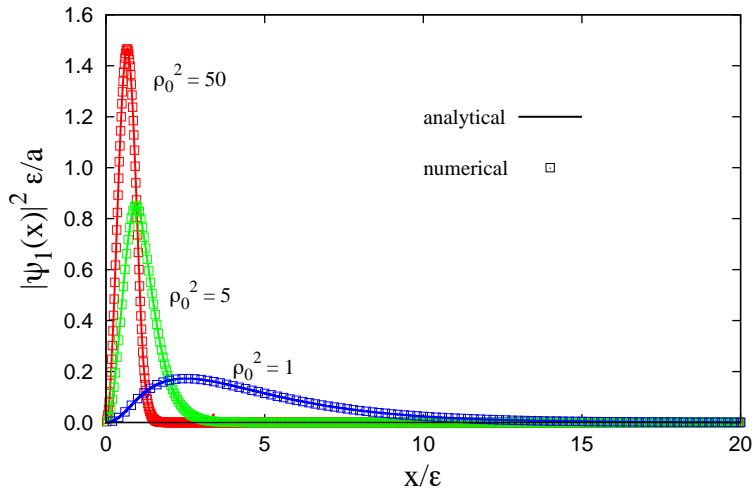}
 \caption
{ 
The probability, multiplied by $\epsilon/a$, vs. $x/\epsilon$ for three different strengths, $\rho_0^2 = 50$,~$5$~and~$1$. Both numerical and analytical results
are shown. For the numerical results we used $\epsilon/a = 0.001$ to ensure that the result for $\rho_0^2 = 1$ was oblivious to the wall of the infinite square well. 
However, they are universal (with respect to $\epsilon$) and clearly show how the wave function becomes more concentrated at the origin as the strength increases.
For each case taking $\epsilon \rightarrow 0$ results in infinitely bound energies and collapsed wave functions.
} 
\label{fig8}
\end{figure}

To illustrate this scaling we first re-plot results from Fig.~\ref{fig5}, but now we plot the probability, $|\psi_1(x)|^2$, multiplied by $\epsilon$, vs. $x/\epsilon$ (not $x/a$)
in Fig.\ref{fig7}. This is how we first realized this scaling [even though it is obvious from Eq.~(\ref{wavefn})].  We also see that
the value of $\epsilon/a$ need not be too small, but this of course depends on the value of $\rho_0^2$.
The exact analytical result, given by squaring Eq.~(\ref{wavefn}), is also shown with a black curve and of course agrees with the numerical result.
%figure 7
This function is universal in that it does not depend on $\epsilon$.
Of course as $\epsilon \rightarrow 0$ the wave function itself would become non-normalizable, and the length scale (in $x$) of the non-zero amplitude of the wave function
would collapse to the origin, but this result tells us what the result looks like as this limiting process is taken. For instance, wild oscillations do not occur, 
and everything is well-behaved. Finally, in Fig.~\ref{fig8} we show the same graph, but for several progressively smaller values of potential strength, $\rho_0^2$.
%figure 8
It is clear that identical results are obtained through the numerical and analytical methods, and, as expected, as $\rho_0^2 \rightarrow 1/4$ the singular wave function
becomes more extended. Nonetheless, these are universal functions, and do not depend on $\epsilon$ except through the axis labels, even though an actual value of
$\epsilon/a$ was required for the numerical method. Similar results and agreement can be shown for the excited states.

\section{SUMMARY}

We have carried out a study of an attractive single particle potential, $-\alpha/x^2$ for $x > 0$, known to show extreme anomalous properties. 
While several studies have examined this potential before us, we have done two things in addition: (i) we have adopted a somewhat different 
regularization procedure and (ii) we have provided a complementary procedure for solution, through a matrix mechanics approach previously used 
for many other one-body potentials.  The former approach suffers from the need to utilize Bessel functions with imaginary index, for which we used both
established subroutines (in Maple) and ones we wrote ourselves (in Fortran). Either way, these are not so familiar to undergraduates (or almost anybody else!),
so the secondary approach, while ``numerical,'' allows a more ``hands-on'' approach for undergraduates, and therefore provides some extra freedom for experimentation.
Indeed, after the calculations for this problem were completed, we first became aware of the newest (3rd) edition of a very popular textbook 
on Quantum Mechanics,\cite{griffiths18} where a study of this potential was included as a problem (Problem 2.60). We would recommend a complementary 
study of the same potential with the matrix mechanics approach explained in this paper and previous references (which differs significantly from the matrix approach suggested in Problem 2.61 of the same Ref.~[\onlinecite{griffiths18}].

In particular, we feel that two lessons were achieved that are valuable for the reader (and for ourselves). First, insights not so forthcoming with unfamiliar 
non-elementary functions can be achieved with an alternative (and simpler) approach. Matrix mechanics requires only a first year knowledge of integral calculus and
of linear algebra (plus an ability to use software that calls a diagonalization routine.\cite{randles19}) Secondly, it is always desirable to have two completely 
independent methods of solution for any problem. While this is not always achievable for all problems, it is here, and 
particularly for the novice, is almost crucial to build the confidence that a correct and accurate solution has been attained.

One cannot really solve for the ground state of the $-\alpha/x^2$ potential (with $2m_0 \alpha/\hbar^2 > 1/4$); however, with the cutoff near the origin introduced here
the problem is readily solved, and shows all the usual characteristics of an attractive potential in one dimension. We have shown how one can use the numerical matrix mechanics, with the simplest of bases, to successfully obtain accurate numerical results for the low-lying levels for the regularized form of the pure $-\alpha/x^2$ potential.
Instead of advanced knowledge about the modified Bessel function with imaginary index and self-adjoint extensions, the mathematics required to solve the
problem numerically with the regularized potential is minimal.
This numerical skill set, though rare a generation ago, is becoming increasingly useful and common among physics students at the undergraduate level and beyond.

\begin{acknowledgments}

This work was supported in part by the Natural Sciences and Engineering
Research Council of Canada (NSERC). We especially want to acknowledge the support of the Canada-ASEAN Scholarships and Educational
Exchanges for Development (SEED) program, that allowed one of us (TXN) to visit the University of Alberta for an eight month period.

\end{acknowledgments}


\begin{thebibliography}{9}
\bibitem{remark} We have used two independent determinations, one with Maple, and one with our own subroutines for Bessel functions
with imaginary index written in Fortran.
\end{thebibliography}
\end{document}